\documentclass[aps,reprint,twocolumn,amsmath,amssymb]{revtex4-1}
\pdfoutput=1
\usepackage{graphicx}

\begin{document}
\title{A roadmap for searching cosmic rays correlated with the extraterrestrial neutrinos seen at IceCube}

\author{J.A. Carpio and A.M. Gago}
\address{Secci\'on F\'isica, Departamento de Ciencias, Pontificia Universidad Cat\'olica del Per\'u, Apartado 1761, Lima, Per\'u}

\begin{abstract}
We have built Sky maps showing the expected arrival directions of 120 EeV ultrahigh energy cosmic rays (UHECR) directionally correlated with 
the latest astrophysical neutrino tracks observed at IceCube, including the 4-year high-energy starting events (HESE) and the 2-year 
Northern tracks, taken as point sources.We have considered contributions to UHECR deflections from the galactic and the extragalactic magnetic 
field, and a UHECR composition compatible with the current expectations. We have used the Jansson-Farrar JF12 model for the Galactic magnetic 
field and an extragalactic magnetic field strength of 1nG and coherence length of 1Mpc. We observe that the regions outside of the Galactic 
plane are more strongly correlated with the neutrino tracks than those adjacent to or in it, where IceCube HESE events 37 and 47 and, from
the 2-year Northern hemisphere sample, event $N13$ are good candidates to search for excesses, or anisotropies, in the UHECR flux.
On the other hand, clustered Northern tracks around $(l,b)=(0^\circ,-30^\circ)$ are promising candidates for a stacked point source search. As 
an example, we have focused on the region of 150 EeV UHECR arrival directions correlated with IceCube HESE event 37  located at 
$(l,b)=(-137.1^\circ,65.8^\circ)$ in the Northern Hemisphere, far away from the Galactic plane, obtaining an angular size $\sim 5^\circ$, 
being $\sim 3^\circ$ for 200 EeV, and $\sim 8^\circ$ for 120 EeV.
\end{abstract}

\maketitle
\flushbottom

\section{Introduction}

TPhe discovery of extraterrestrial neutrinos made by the IceCube (IC) Neutrino Observatory \cite{Aartsen13,Aartsen14,Aartsen15} has 
boosted the multimessenger searches of point sources, which eventual results should lead us to understand the high-energy 
astrophysical phenomena. Within this context, it is believed that extraterrestrial neutrinos are created 
inside or outside the source, primarily through 
photopion production. The pion production is caused by the interaction of  
ultrahigh energy cosmic rays (UHECR) either with the cosmic microwave background (CMB) or 
with the extragalactic background light (EBL) \cite{Aloisio16}, yielding neutrinos from 
their decay, with energies in the 
range of 10PeV-1EeV or in $\mathcal{O}$(PeV), respectively. Thus, given the connection
between UHECR and neutrinos, some degree of correlation between  
their respective experimental observations is expected. This kind of study has been already conducted 
using the extraterrestrial neutrinos observed at IceCube and 
a combined UHECR data from the Pierre Auger Observatory(PAO) and Telescope 
Array (TA), with not a positive outcome yet \cite{Aartsen16}. There have also been other 
attempts to seek correlations between photons and neutrinos 
\cite{Santander15} or gravitational waves with neutrinos \cite{ICLIGOVIRGO}. In the 
future, multimessenger searches, such as joint neutrino/gamma-ray transient sources,
will be facilitated by AMON \cite{Smith13}.

The correlation analysis between UHECRs and neutrinos, as has been done in \cite{Aartsen16}, relies 
on the distribution of cosmic ray arrival directions. This paper uses another approach 
to this issue. In our case, we will focus on predicting the regions on the Sky 
where UHECRs correlated with neutrinos are expected to arrive, considering 
that the neutrino tracks are pointing 
to the sources. In this way, these regions will 
constitute a tool for searching UHECR excesses on the Sky. Besides, searches 
in these regions could be used as complementary test of the various hypothesis implied 
in their construction, among others, the magnetic field model, the UHECR 
composition and, at a more fundamental level, the expected 
associated production of UHECR and neutrinos. 

In fact, the choice of the Galactic and extragalactic magnetic field model is one of 
the most important hypothesis in our work. These fields deviate the 
UHECRs from its path to the Earth, making 
their arrival directions to not coincide with the 
corresponding ones of the neutrinos. For the extragalactic 
magnetic field (EGMF), we will use a turbulent field of strengths $\sim 1$nG 
and coherence lengths $\gtrsim 1$ Mpc 
following the references \cite{Dolag04,Sigl05}. For the 
Galactic magnetic field (GMF), we will use field strengths of $\sim 1 \mu$G. The GMF 
is divided into a regular and a turbulent component, 
the former described by models such as those in \cite{Pshirkov11,Jansson12a} and 
the latter in \cite{Jansson12b}. The GMF deflections are 
dominated by the regular components, to which is added, as a secondary effect, a
smearing due to the turbulent component. Another premise in the calculation 
of the magnetic deviation is the UHECR mass 
composition, which is taken into account
in this paper, as it is described in sections ahead. Currently, 
the PAO has yet to explore 
the mass composition above 50 EeV, although a trend 
towards a heavy composition above 10 EeV is apparent \cite{Aab14,Porcelli15}. 
We select 13 muons tracks from the extraterrestrial neutrino data sample given by IceCube in the 79-string and 86-string 
configurations \cite{Aartsen13,Aartsen14,Aartsen15}, which spans 
the deposited energy range 60TeV-1PeV. This is equivalent to four years of data taking and gives us an $E_\nu^{-2.58}$ 
neutrino flux spectrum and a flux of $2.2\pm 0.7\times 10^{-18}$ GeV$^{-1}$cm$^{-2}$s$^{-1}$sr$^{-1}$ at 100 TeV. We choose 21
muon neutrino tracks from the two-year sample \cite{Aartsen15b}, consisting of tracks coming from the Northern Hemisphere, containing
approximately 35000 muon tracks. A subsample of the 21 tracks with the highest energies was released and are likely to be of 
astrophysical origin. \footnote{This subsample is found in https://icecube.wisc.edu/science/data/HE\_NuMu\_diffuse}

The paper is divided as follows: in section II we describe
the analysis ingredients, which are: the extragalactic magnetic field 
deflections with its corresponding treatment for UHECR propagation, the  
galactic magnetic field deflection and the definitions 
for signal and background. In section III we present our results
and, finally, in section IV our conclusions. 

\section{Analysis ingredients}
We subdivided this section into three parts: the EGMF deflections 
and UHECR propagation, GMF deflections and the Signal and Background definitions.

\subsection{EGMF Deflections and UHECR propagation}
Typical deflections in a turbulent EGMF with a Kolmogorov spectrum 
are given by \cite{Sigl05}
\begin{equation}
\delta_\textrm{rms} = 0.8^\circ Z \left(\frac{B_0}{E}\frac{10^{20} \textrm{eV}}{10^{-9}\textrm{G}}\right)\sqrt{\frac{D}{10\textrm{Mpc}}}
\sqrt{\frac{L_c}{1\textrm{Mpc}}},
\label{EGMFTypical}
\end{equation}
where $B_0$ is the EGMF 
root-mean-square 
field strength, $E$ is the UHECR energy, $L_c$ the coherence length of the field and $D$ is 
the propagation distance, which starts from the UHECR. There is no 
general consensus on the values of $B_0$ and $L_c$ \cite{Dolag04,Sigl05}, in particular,
we are using the values of 1 nG and 1 Mpc, respectively. Due to the large propagation 
distances of order $>10$ Mpc, energy losses are 
taken into account being obtained from forward tracking via Monte Carlo simulation using CRPropa 3\cite{Batista13}. These energy loss 
processes include: cosmological expansion, photopion production and photodisintegration. For the latter two processes provide both the CMB 
and the infrared background light (IBL) described in \cite{Gilmore12}.

We estimate the magnetic deflections 
through the injection of individual events from the following spectrum \cite{Roulet13}
\begin{equation}
Q_Z(E_p) \propto \frac{E_p^{-\gamma}}{\text{cosh}[E_p/(ZR_\text{max})]},
\end{equation}
where $E_p$ stands for primary UHECR energy 
and $R_\text{max}=20$ EV marks 
the rigidity cutoff, where rigidity 
is defined, in general, as $R=E/(Ze)$. The sources emit p,He,N,Si,Fe nuclei
\cite{Roulet13} according to the following ratios
\begin{equation}
\text{p:He:N:Si:Fe} = 0.1:0.27:0.30:0.32:0.005
\label{OurComposition}
\end{equation}
and we assume a homogenous distribution of identical sources. The composition in Equation \eqref{OurComposition} fits well the Auger 
data reasonably and gives us a maximum distance of $\sim$200 Mpc from which UHECR above 100EeV may reach the Earth. The propagation distance 
$D$ decreases exponentially with the UHECR arrival energy $E$. 

\begin{figure}
\centering
\includegraphics[width=0.5\textwidth]{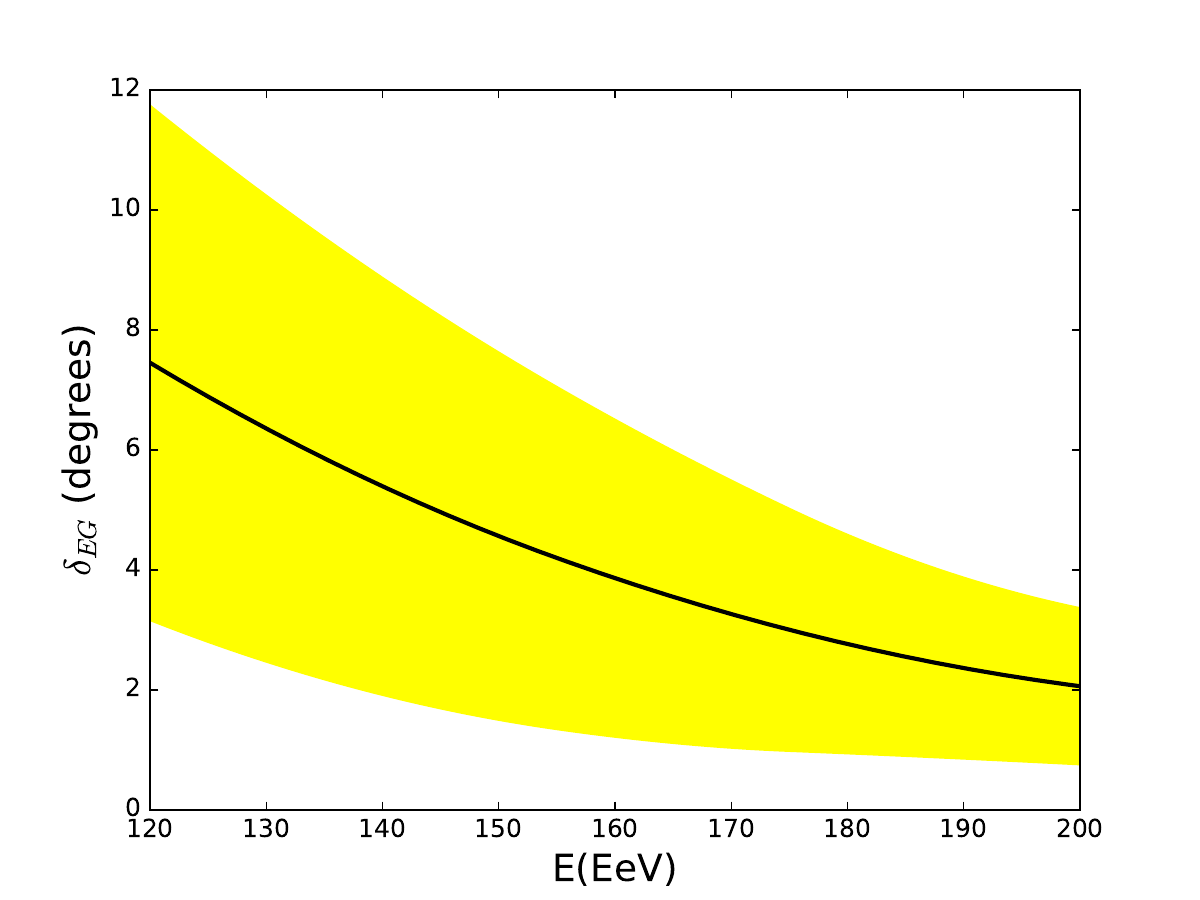}
\caption{$\delta_{EG}$ as a function of its arrival energy.}
\label{EGDeflectionsFig}
\vspace{-10pt}
\end{figure}

We then generate $10^6$ Monte Carlo events, calculating the magnetic deflection 
with Equation \eqref{EGMFTypical} for small 
steps $\Delta L$, due to the energy losses, adding 
them in quadrature. This amounts to the substitution
\begin{equation}
\frac{Z}{E}\sqrt{D} \longrightarrow \sqrt{\sum_{i=1}^N \frac{Z^2(L)}{E^2(L)} \Delta L},
\label{EGMFEffective}
\end{equation}
where $N\Delta L=D$. These magnetic deflections also increase the UHECR propagation length by $\Delta D=\sum_{i=1}^N \Delta r$ where
\begin{equation}
\Delta r \simeq 0.195\text{Mpc} \;\frac{Z^2}{(E/\text{EeV})^2}\frac{L_c}{1\text{Mpc}} \left(\frac{\Delta L}{1\text{Mpc}}\right)^2.
\end{equation}

Therefore, we obtain the deflections using \eqref{EGMFTypical} and \eqref{EGMFEffective} for a distance $D$, we let the particle 
propagate an additional $\Delta D$, to see if the particle loses any extra energy. We use the latter approach, looking at extra energy
losses, because $\Delta D\ll D$ and does not significantly contribute to the deflection

Finally, once the particle enters the Galaxy with an arrival energy $E$, we assign a deflection $\delta_\textrm{EG}$ per energy bin 
of width $\Delta E$ via
\begin{equation}
\delta_\textrm{EG}(E) = \textrm{Average } \delta_\textrm{rms} \textrm{ in bin},
\label{EGMFMean}
\end{equation}
where $E\in [E_i,E_i+\Delta E]$. In figure \ref{EGDeflectionsFig}, we display the values of $\delta_\textrm{EG}$ as a function of E. A band 
$[\delta_\textrm{EG}-\omega,\delta_\textrm{EG}+\omega]$ is included, such that 68\% of 
the events in the bin are enclosed in this interval. Where we start with 
deflections intervals ranging between $[3^\circ,12^\circ]$ at 120 EeV until reach deflections intervals, as 
small as $[1^\circ,3^\circ]$ at 200 EeV.     

\subsection{GMF deflections}
Once the cosmic ray enters the Galaxy with an energy $E$, we can ignore energy loss processes due to the relatively small size of the Galaxy 
$\sim 40$ kpc and trajectory lengths $<100$ kpc. Thus, the UHECR arrival energy at Earth 
is also $E$ and the Galactic magnetic deflections
will depend on the rigidity $R=E/(Ze)$.

We use the JF12 GMF model \cite{Jansson12a,Jansson12b}, which is designed to fit the WMAP7 Galactic synchrotron emission map and more than
40,000 extragalactic Faraday rotation measurements. It is divided into three components: regular (coherent), striated (anisotropic) and 
an isotropic random field. The latter two are small-scale fields and will be referred to as the random or non-coherent components. 
We use the best fit values in \cite{Jansson12a}, for the coherent field parameters, and in \cite{Jansson12b}, for the random field parameters.
The small scale field is assumed to have a coherence length of 60 pc. 

We use the backtracking method to determine GMF deflections. For a hypothetical cosmic ray arriving at Earth, we reverse its
incoming momentum vector, then change the sign of its charge. We propagate this particle from the Earth, through the GMF, until it
leaves the Galaxy. The backtracking is performed using the Runge-Kutta methods in CRPropa 3.

\begin{figure}
\centering
\includegraphics[width=0.5\textwidth]{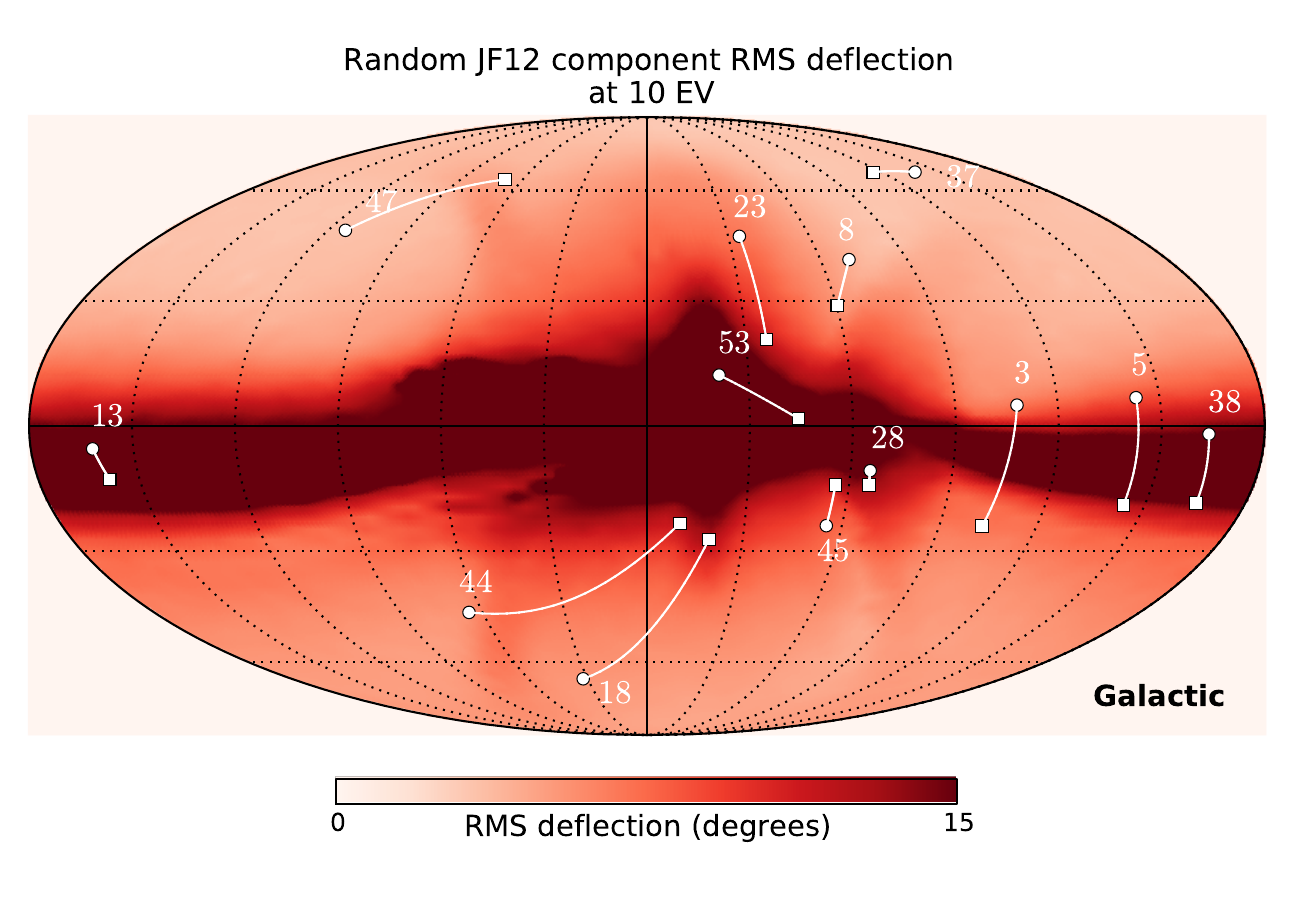}
\caption{RMS deflection of cosmic rays with rigidity 10EV. The white circles correspond to the reconstructed directions of HESE neutrino 
tracks and the white squares mark the expected arrival direction assuming only the JF12 coherent field. The white lines joining both are to 
match UHECRs with their corresponding track, they do not show the actual trajectory taken by the particle. Tracks are labeled by event numbers 
given in \cite{Aartsen15}.}
\label{JF12RMS}
\end{figure}

We treat the GMF coherent and turbulent components separately, taking advantage of the JF12 Field parametrization outlined in 
\cite{Carpio15}. Firstly, for a given initial direction $P_0=(l_0,b_0)$ of a UHECR 
with energy $E$, where $(l,b)$ are in the galactic coordinate system with $-180^\circ\leq l<180^\circ$ the galactic longitude and 
$-90^\circ\leq b\leq 90^\circ$ the galactic latitude, we backtrack it through the coherent field to the position $P_c=(l_c,b_c)$. 
Starting from $P_c$, we perform an additional non-coherent field deflection, given by the von Mises-Fisher distribution 
\cite{Fisher53}
\begin{equation}
\label{JF12DefFormula}
f (\delta) = \frac{\kappa}{e^\kappa-e^{-\kappa}} \exp\left(\kappa\cos\delta\right),
\end{equation}
where $\kappa= \kappa(l_0,b_0,R)$ is a fit parameter and $\delta$ is the angular distance
between $P_c$ and the final position of the particle. We also assume that the azimuthal 
distribution of this random deflection is flat. Contrary to the parametrization in \cite{Carpio15}, we were forced to extend the Rayleigh 
distribution to a von Mises-Fisher distribution in order to handle deflections that 
are not so small. This is caused by the low rigidity particles. 
The energy dependence in $\kappa$ is given by
\begin{equation}
\kappa(l_0,b_0,R) = A_1(l_0,b_0) R+ A_2(l_0,b_0) R^2.
\label{KappaFit}
\end{equation}
This approximation has been tested in the rigidity range $10 \text{EV}\leq R \leq 100 \text{EV}$. The parameters $A_1,A_2$ were obtained
using HEALPix \cite{Gorski05} to divide the sky into 3072 pixels of equal solid angle. We emphasize 
that in the vicinity of the Galactic
plane, where large deflections are present provided by the high
turbulent fields components, the parametrization given in 
Equation \eqref{KappaFit} is unreliable and we solve these cases numerically. 

In the small deflection hypothesis (valid for $<15^\circ$), 
where $\kappa\gg 1$, concentration parameter $\kappa$ is related to the root-mean-square deflection 
\begin{equation}
\delta_\textrm{Gal} = \frac{1}{\sqrt{\kappa}}.
\label{DeltaGal}
\end{equation}

We assume an average rigidity $\langle R \rangle_E $ for all particles with a given energy $E$, which obeys the relation 
$\langle R\rangle_E \approx (E/10.5$EeV)EV according to our simulations.  

The root-mean-square deflections at $R=10$ EV for different arrival directions are shown in figure \ref{JF12RMS}. We see that trajectories
close to the Galactic center and/or plane can be affected by high ($>15^\circ$) non-coherent deflections which exceed the angular resolution
of experiments ($\sim 2^\circ$) by an order of magnitude. As a reference, we have included the 
reconstructed arrival directions of the high-energy starting events (HESE) neutrino tracks, labeled according to their corresponding event
numbers as presented in \cite{Aartsen15}. We also marked the respective arrival directions of 10 EV UHECR, 
considering the aforementioned tracks
as point source and ignoring EGMFs and the JF12 random field components.

\subsection{Signal and Background}
We work in a similar scenario such as described in \cite{Aartsen16}, where a sample of $N_{CR}$ UHECR and $N_\nu$ neutrinos is given. The
neutrinos are considered as point sources, while the $N_{CR}$ cosmic rays are a combination of signal and background events. 
We define $\mathcal{S}_i^j$ as the probability density (pdf) that the ith cosmic ray came from the direction of the jth neutrino event.
\begin{equation}
\mathcal{S}_{i}^j = \frac{\kappa_i}{2\pi(e^{\kappa_i}-e^{-\kappa_i})}\textrm{exp}(\kappa_i \mathbf{x}_i\cdot\mathbf{x}_j),
\end{equation}
\begin{figure*}[t]
\centerline{%
\includegraphics[width=0.40\textwidth]{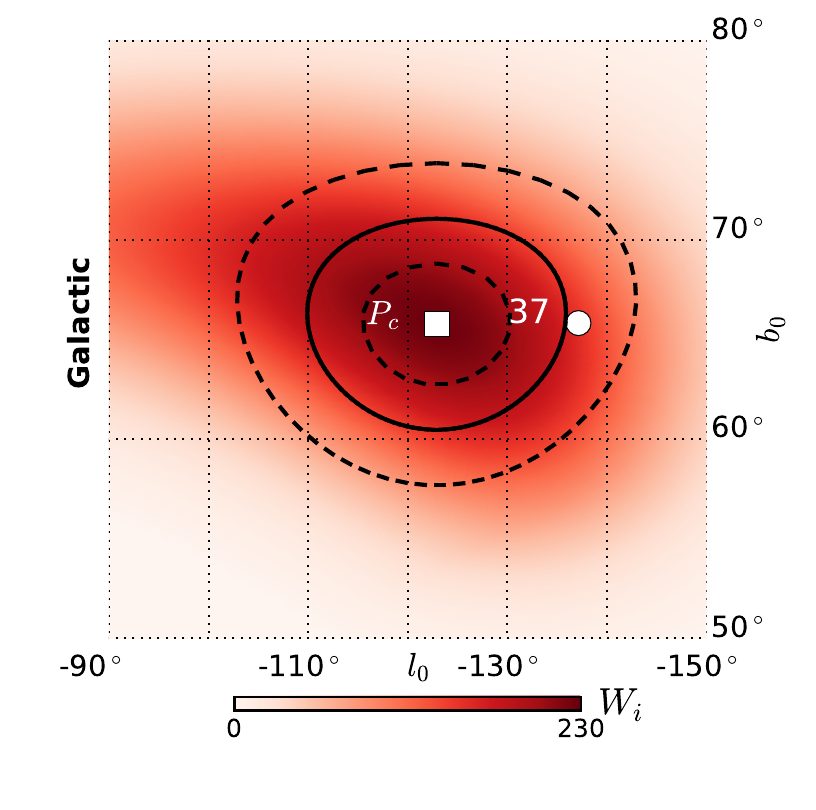}
\includegraphics[width=0.55\textwidth]{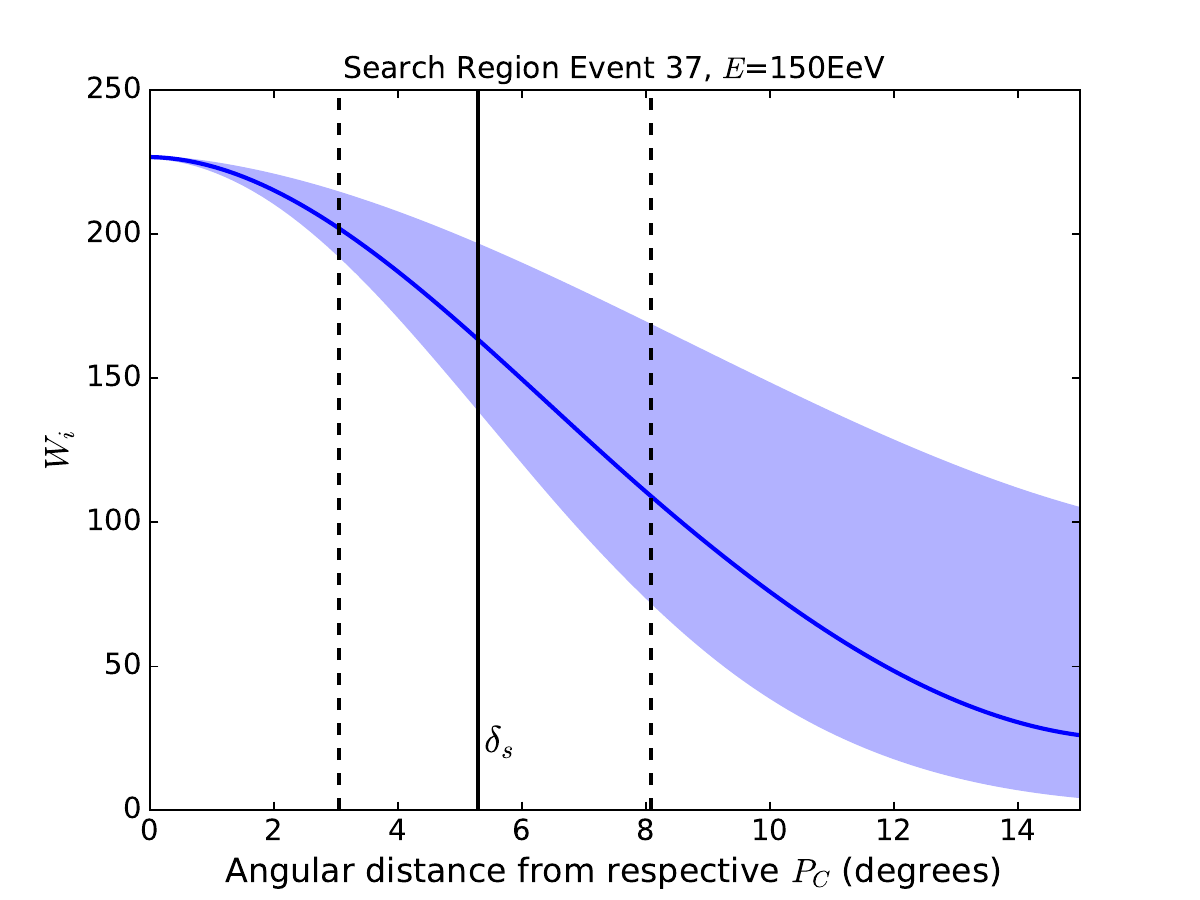}
}%
\vspace{-5pt}
\caption{$W_i$ for a single point source search located at Event 37 and 150EeV UHECR. At the 
left we have the corresponding two-dimensional plot of $W_i$ in 
the coordinates $l_0$ and $b_0$. At the right we have the one-dimensional projection of 
$W_i$ in the angular distance respect to $P_c$}
\label{Track37-150}
\end{figure*}
where $\kappa_i = 1/\sigma_i^2$ and $\sigma_i$ accounts for the overall smearing of the ith cosmic ray. In the 
limit of small smearings $\mathcal{S}_i^j$
reduces to a two-dimensional Gaussian. For single source searches, when $N_\nu=1$, the signal pdf is $\mathcal{S}_i\equiv \mathcal{S}_i^1$ 
and is used in the unbinned likelihood 
analysis like the mentioned in \cite{Braun08}. For the so-called stacked source searches, 
when $N_\nu>1$, we add up the contributions
from multiple faint sources and the signal pdf is modified to
\begin{equation}
\mathcal{S}_i = \frac{1}{N_\nu}\sum_{j=1}^{N_\nu} \mathcal{S}_i^j.
\label{DefinitionSiStacked}
\end{equation}

In order to use any of these formulas, we substitute the arrival direction $\mathbf{x}_i$ of the UHECR by its backtracked direction 
$\mathbf{x}'_i$, assuming that the only magnetic field involved is the regular JF12 component and that the particle's rigidity is given 
by $\langle R\rangle_E$. We then determine the values of $\kappa_i$ and $\delta_\textrm{Gal}$ via 
Equations \eqref{KappaFit}, that parametrizes 
the smearing effect of the non-regular component,and 
\eqref{DeltaGal}, respectively, which are 
functions of the UHECR energy and its arrival direction. The EGMF deflections and angular
resolution effects are incorporated by 
making the substitution
\begin{equation}
\kappa_i \longrightarrow \frac{1}{\sqrt{\delta_\textrm{Gal}^2 + \delta_{EG}^2+\delta_\textrm{res}^2}},
\label{CombineRMS}
\end{equation}
with $\delta_\textrm{Gal}=\delta_\textrm{Gal}(\langle R\rangle_E)$, $\delta_{EG}=\delta_{EG}(E)$
and $\delta_\textrm{res}$ is the angular resolution of the experiment. We will assume an energy 
independent deflection $\delta_\textrm{res}=2^\circ$ as a characteristic angular resolution for IceCube 
and UHECR ground array experiments.

For the large GMF deflections present at the Galactic plane, where Equation \eqref{KappaFit} 
is not valid, $\mathcal{S}_i$ is determined entirely via Monte Carlo. 
We also define $\mathcal{B}_i$ as the probability density that the CR is a background event. Typically $\mathcal{B}_i=1/4\pi$
or, in the case of the analysis in \cite{Aartsen16}, the normalized exposure of the experiment. 

\section{Results}

Now, we quantify how likely an observed UHECR, with a given arrival direction $\mathbf{x}_i$ and 
energy $E_i$, could have the same origin as the  neutrino track, which we treat as the UHECR point source, through 
the following ratio: 
\begin{equation}
W_i = \dfrac{\int \mathcal{S}_i d\Omega}{\int\mathcal{B}_i d\Omega},
\end{equation}
where $d\Omega$ is integrated in a region of $1^\circ$ in the Sky, centered around the neutrino track. 

\begin{figure}[t!]
\centering
\includegraphics[width=0.5\textwidth]{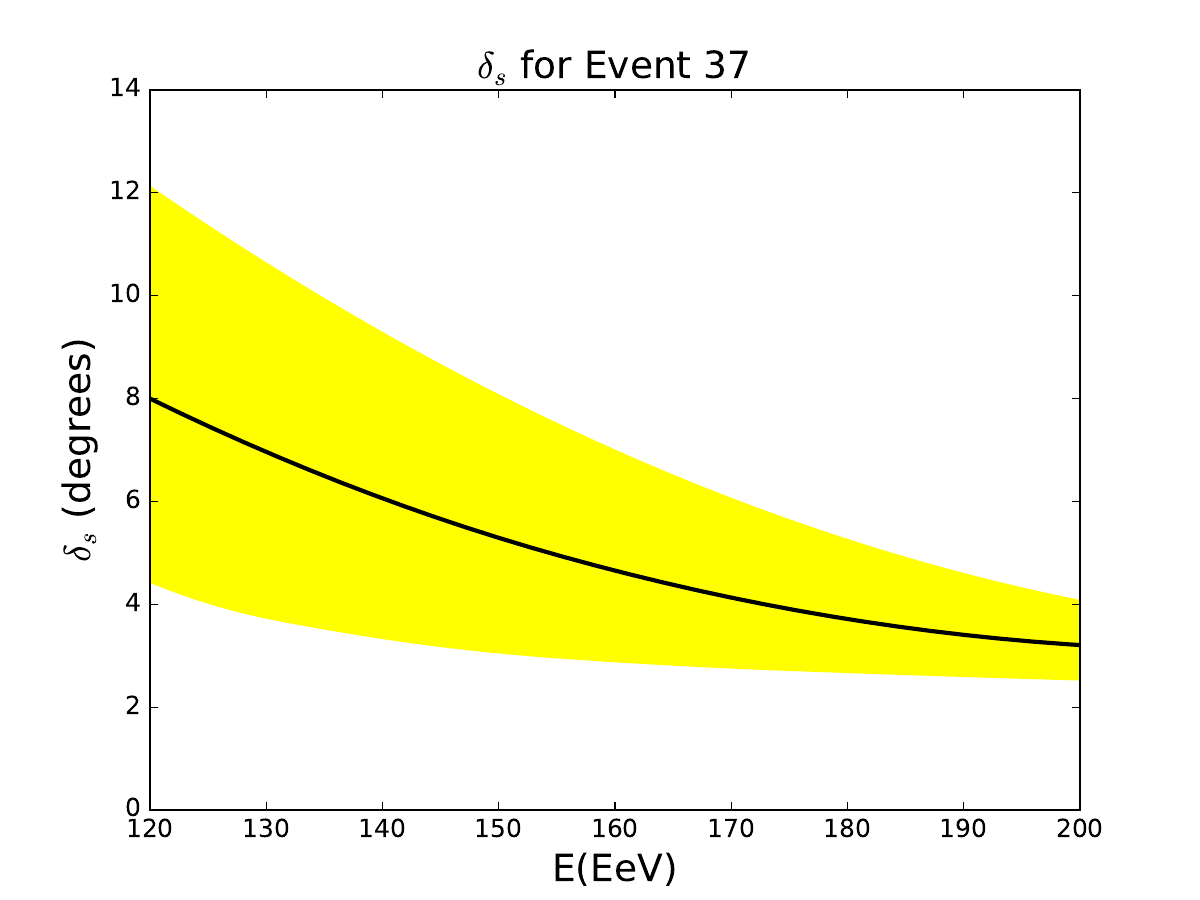}
\caption{$\delta_{s}$ angular size region for the IC event 37 as a function of UHECR arrival energy}
\label{DeltaSFig}
\end{figure}

In figure \ref{Track37-150}, we show the effect of $W_i$ in a single point source search, 
located at IC event 37 $(l,b)=(-137.1^\circ,65.8^\circ)$ in the Northern hemisphere, labeled according 
to the numbering in Reference \cite{Aartsen16}, in two plots. One is a two-dimensional map in the 
coordinates $(l,b)$, where $P_c$ marks the expected UHECR arrival direction, when 
considering only the JF12 coherent component.
The other one is its corresponding one-dimensional projection in angular distance centered in 
$P_c$, where the solid line is the average value of $W_i$ and 
the shaded region covers the whole set of values given by the points at the
angular distance contour. We have selected IC event 37 
because it belongs to a region where the 
GMF random component deflections are very small, as shown in figure \ref{JF12RMS},
being the random EGMF responsible for most of the 
smearing around $P_c$, displayed in the two-dimensional plot. These characteristics turn IC event 37 into
a good 
candidate for searches of UHECR excesses around it. In both plots it is clear that $P_c$
gives the highest ratio, since, by construction, it is here where the signal pdf is 
maximized. Ellipsoidal(vertical) solid/dashed lines are shown for the two(one) dimensional plot. The 
solid line represents the size of the typical or average angular search region 
$\delta_{s}=\sqrt{\delta_\textrm{Gal}^2 + \delta_\textrm{EG}^2 + \delta_\textrm{res}^2}$, measured 
from $P_c$, while the dashed lines include the effects of the spread in $\delta_\textrm{EG}$ shown in 
figure \ref{EGDeflectionsFig}.
Our results show that UHECRs confined within the 
region of size $\delta_{s}$, that have the appropiate energy and arrival 
direction, would have very high values of $W_i$. 
In the one-dimensional plot, a band enclosing  
the $W_i$ average is also displayed, and is caused by the anisotropy
of the random deflections, or, equivalently, the B field 
itself. Otherwise, if the random deflections were isotropic, there 
would be no band, which means that all the values 
would converge to a single one. For small angular distances, 
the variation, or width of the band, of $W_i$, is small 
because the parameter $\kappa_i$ is essentially constant, and as 
we move away from $P_c$ this variation is significant.
\begin{figure*}[t!]
\centerline{
\includegraphics[width=0.5\textwidth]{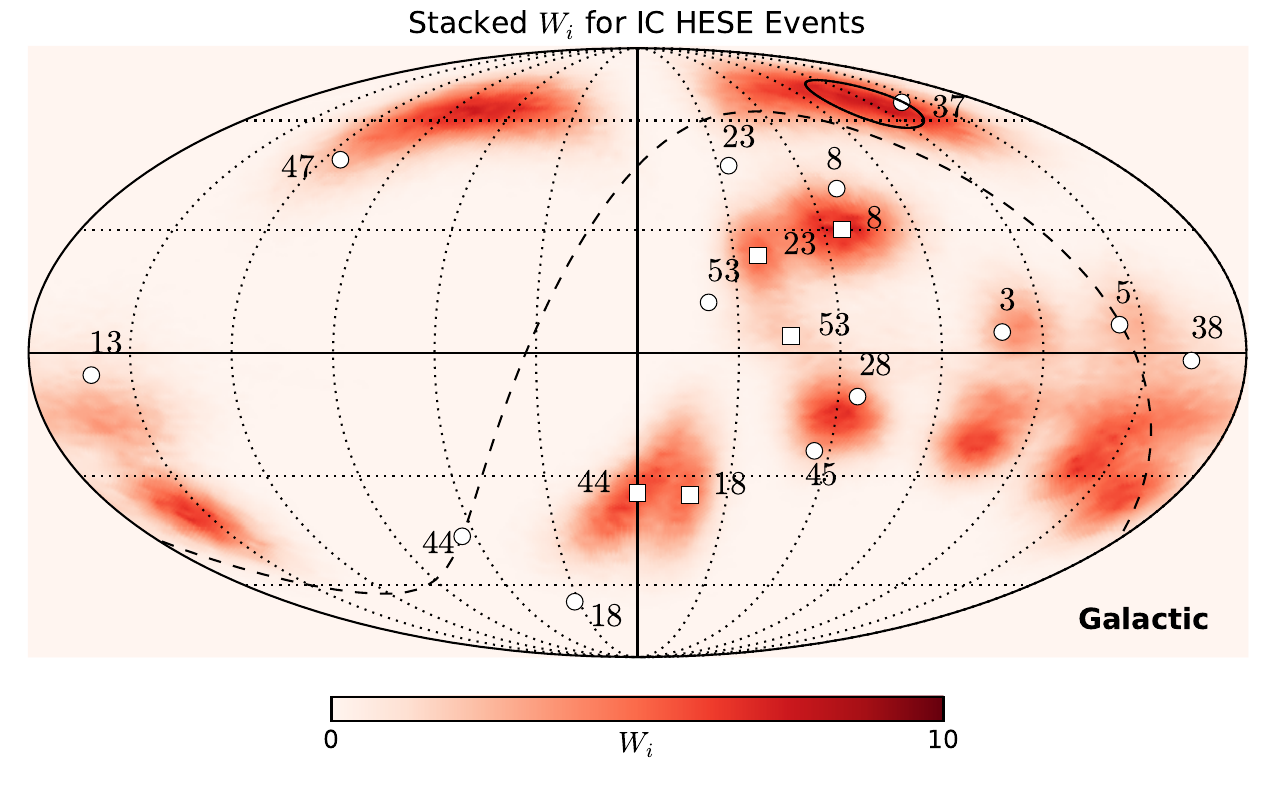}
\includegraphics[width=0.5\textwidth]{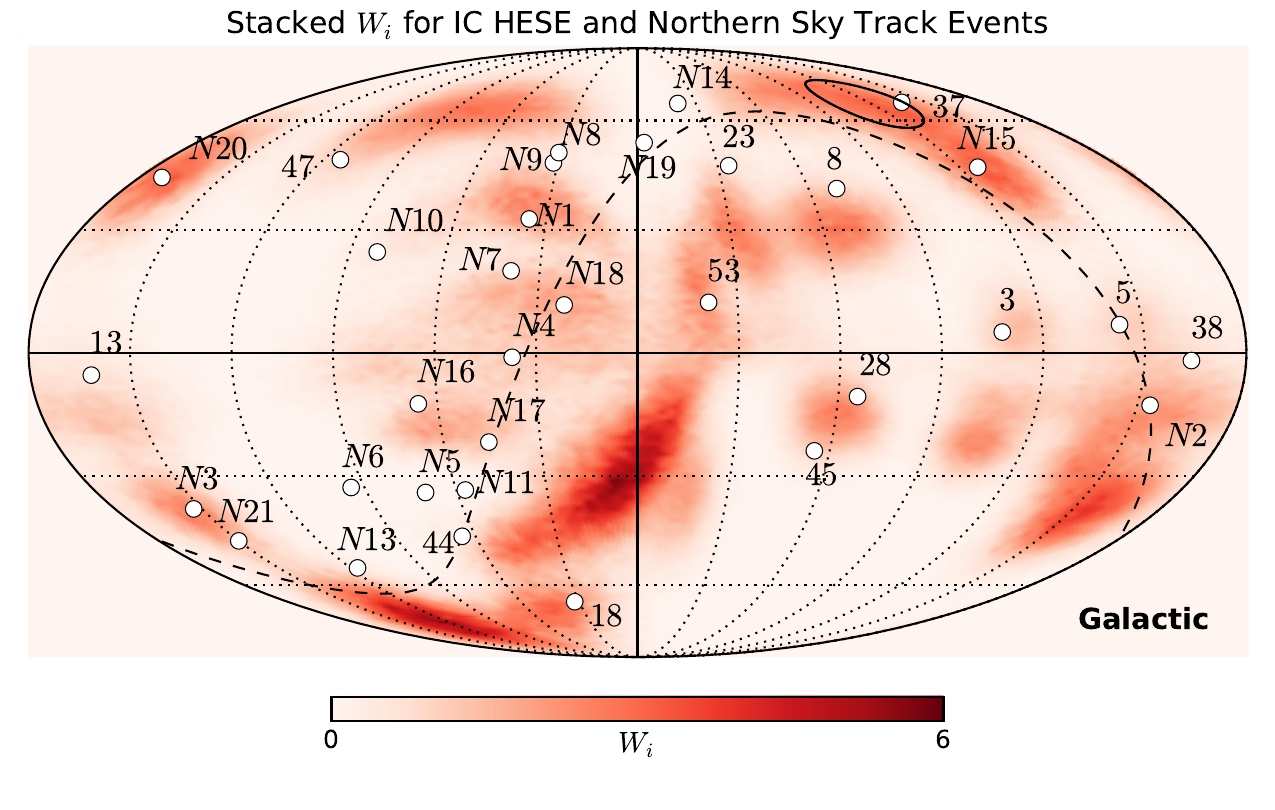}
}
\caption{Sky map for 120 EeV cosmic ray events of undetermined mass, using the 4-year IC HESE events 
(left panel) and a combined sample of IC HESE events and the 2-year IC Northern tracks (right panel).
The white circles mark the neutrino tracks: HESE tracks are labeled according
to their event numbers in \cite{Aartsen15}; Northern tracks, preceded by the letter N, are numbered according to the order in which they 
appear in the IceCube Data Release. The white squares mark the respective $P_c$ of the track.
Event N12 is not included in the map because it coincides with HESE track 5.
The black dashed line marks the Equatorial plane. Suggested search region for event 37 is marked with a solid black line.}
\label{SkyMap120EeV}
\end{figure*}

For a wider perspective, we are showing in figure \ref{DeltaSFig} the dependence of $\delta_s$ with
UHECR arrival energy. It follows the same behaviour of $\delta_{EG}$ (see figure \ref{EGDeflectionsFig})
since $\delta_s$ in the case of IC event 37 is dominated by the extragalactic contribution. 

It is important to mention that in case the current or future ground array experiments were able to
identify the UHECR mass, we would have a sensible improvement in our analysis since the uncertainties
in the rigidity (i.e. $\delta_\textrm{EG}$) are going to be small and inside the Galaxy we would know
with greater precision the backtrack trajectory because $R$ is constant within the Galaxy. The analysis
is also dependent on the UHECR-neutrino sources being within the GZK sphere, such that
we may observe the cosmic rays.

The sky map of $W_i$ for a stacked search of cosmic ray events by using the IceCube tracks as
the assumed point sources is presented in figure \ref{SkyMap120EeV}, which assumes incoming 
120EeV cosmic rays, which is our lower energy limit in our analysis, and unknown mass number. 
This figure contains two Sky maps, one for the IC HESE event sample \cite{Aartsen16} and another
for the combined sample of HESE events and the second one using both the 4-year HESE tracks and the 
2-year IceCube Northern Sky tracks\cite{Aartsen15b} in a combined search.
As before, the circles represent the positions of the neutrino tracks. In the HESE Sky map,
we also included squares indicating their respective
points $P_c$ when the neutrino tracks are close to each other. 
Here we observe that for events close to the Galactic plane, we do not have a 
well-defined region to correlate with the neutrino track. Instead, we have disjoint regions,
above and below the Galactic plane, which represent similar low values
of $W_i$, such as those in neutrino events 3, 5 and 38. 
This is caused by the 
large random component deflections in these regions, as we have seen in figure \ref{JF12RMS}.
There are out-of-plane events, such as events 8 and 23, 18 and 44,
28 and 45, which exhibit well-defined regions,
but with the drawback that they may overlap with each other. Despite our inability to disentangle
the point source of origin, these zones are interesting to look for UHECR excesses, because the effect
of GMF deflections is significantly reduced and obtain higher values of $W_i$.
The ideal case for correlating UHECR with these neutrino tracks are the events 37 and 47, which are 
far away from the Galactic plane. 

In the combined Sky map, we increase the number of regions where we can expect the arrival of UHECRs.
However, due to the nature of the stacked search, which includes a $1/N_\nu$ factor in Equation 
\eqref{DefinitionSiStacked}, the $W_i$ of well-spread tracks is flattened to very low values (see for
example events 8 and 47). Meanwhile, when we have clustered tracks, the $W_i$ are strongly enhanced, as 
we can see in the region close to $(l,b)= (0^\circ,-30^\circ)$. Even when using the enlarged sample, the 
suggested search region for HESE event 37 does not overlap with the others. It is interesting to 
highlight that the Northern event $N13$ exhibits a region with very large values of $W_i$, becoming
a suitable candidate to be included in a point source search.

\section{Summary and Conclusions}

We have built two Sky maps showing different regions where 120 EeV UHECR excesses with respect to the isotropic background should appear:
one for the IC 4-year HESE tracks and another for a combined search using the 4-year HESE events and the 2-year Northern Sky tracks.
These excesses are inferred from the measurement of the correlation between a given UHECR arrival direction with the IceCube neutrino tracks, 
which are taken as point sources. The GMF and EGMF deflections have been calculated, using, correspondingly, the JF12 model and EGMF of
strength $\sim 1$nG and coherence lengths ∼ $\gtrsim 1$Mpc. We note that the out-of-plane regions concentrate higher correlation values, 
quantified by the probability ratio $W_i$, being more promising than the ones near the Galactic plane for revealing excesses. 
Some of these regions can be correlated clearly with a single neutrino track, for instance in events 37,47 and $N13$. These
events are candidates to include in a point source search. For the stacked source search, good candidates include the tracks that
contribute to the region in $(l,b)= (0^\circ,-30^\circ)$.

In particular, we
take a closer look into event 37, where the GMF random component is negligible, considering an energy of 150EeV
and getting a region, where most of the UHECR excess should be located, of angular size $\sim 5^\circ$. If the UHECRs
had energies of 120 EeV or 200 EeV the angular size would be $\sim 8^\circ$ or $\sim 3^\circ$, respectively. This similar
tendency can be extrapolated to the Sky map where we expect regions with a much smaller angular spread, as long as
we increase the UHECR energy. It is clear that the Sky map presented relies on the current statistic of the neutrino tracks and as new data is
released, we may obtain more favourable search regions. Naturally, the possibility to find UHECR in these regions intrinsically depends on
these sources being located within the GZK sphere. Finally, we must say that the results of this analysis should improve if
the ground array experiments identify the UHECR mass.



\acknowledgments

The authors gratefully acknowledge DGI-PUCP for financial support under Grant No. 2014-0064, as well as CONCYTEC for graduate fellowship
under Grant No. 012-2013-FONDECYT. The authors also want to thank Mauricio Bustamante and Jos\'e Bellido for useful suggestions.

\clearpage

\end{document}